\documentstyle[prl,aps,preprint,epsf]{revtex} 
\large

\begin{document}
\draft \title{Dynamical scaling in dissipative Burgers turbulence}
\author{T. J. Newman } \address{Department of Theoretical Physics,\\ 
  University of Manchester,\\ Manchester, M13 9PL, UK} \maketitle
\begin{abstract}
  An exact asymptotic analysis is performed for the two-point correlation
  function $C({\bf r},t)$ in dissipative Burgers turbulence with
  bounded initial data, in arbitrary spatial dimension $d$. Contrary
  to the usual scaling hypothesis of a single dynamic length scale, it
  is found that $C$ contains {\it two} dynamic scales: a diffusive
  scale $l_{D} \sim t^{1/2}$ for very large $r$, and a super-diffusive
  scale $L(t) \sim t^{\alpha}$ for $r \ll l_{D}$, where $\alpha =
  (d+1)/(d+2)$.  The consequences for conventional scaling theory are
  discussed. Finally, some simple scaling arguments are presented
  within the `toy model' of disordered systems theory, which may be
  exactly mapped onto the current problem.
\end{abstract}
\vspace{5mm} \pacs{PACS numbers: 47.10.+g, 05.40.+j, 68.10.Jy}

\newpage

\section{Introduction}

Burgers equation (BE) has found many applications, in both theoretical
and practical senses, over the years since its birth in 1940.  It was
originally proposed\cite{b1} to describe wave propagation in weakly
dissipative media, and, in fact, it is now appreciated\cite{karp} that
within this large class of phenomena there are only two model
descriptions in the limit of weakly non-linear waves; namely BE, and
the Korteweg-de Vries equation\cite{kdv}.  In later years BE was
scrutinised by the turbulence community as a simplified model of
Navier-Stokes turbulence, and thus `Burgulence' was conceived.  The
applications of BE were boosted again in 1986 when Kardar, Parisi and
Zhang proposed that the BE with a stochastic source described the
non-equilibrium evolution of a class of interface models\cite{kpz}.
Under a non-linear transformation, this noisy BE was seen to describe
another rich class of systems, namely directed polymers in random
media which have applications in wetting\cite{abr}, disordered
magnets\cite{huse}, and the pinning of flux lines in
superconductors\cite{blat}.  BE has also received attention as an
approximate model for the formation of large scale structures in the
universe\cite{gurb,lss}.

Naturally, with such a wide range of physical applications, BE has
attracted a great deal of theoretical attention. In the years
subsequent to the revolution in critical phenomena -- when the ideas
of scaling and universality have become so prevalent\cite{amit} --
most theoretical ideas concerning BE are formulated within a `scaling
picture'. Although this is a convenient language for many phenomena,
it must be realised that without a formal renormalization group (RG)
description, scaling must be supported by strong physical insight and
not merely `hand-waving' arguments. As an example, the physics of
domain growth in quenched ferromagnets has been very well understood
on the basis of scaling arguments\cite{bray}, although no explicit RG
calculations have been performed away from the critical temperature.
The domain morphology of this problem provides an excellent basis for
scaling since it is clear that the growing domain scale acts as a
well-defined measure of dynamic correlations (with the caveat that
scalar order parameter domain growth has more subtle scaling due to
the existence of a microscopic scale -- the domain wall thickness).
The concept of dynamical scaling is also supported in this field by a
number of exact calculations (e.g. the one dimensional Glauber
model\cite{glau}, and the large-$n$ limit of the time-dependent
Ginzburg-Landau equations\cite{brnew}), and by a large number of
numerical simulations.

The existence of scaling is not so well established in BE, although
most workers would agree that it is a convenient hypothesis, given the
complexity of the problem. The analytic approach used by
Burgers\cite{b2}, and later by Kida\cite{kida}, certainly demonstrated
the existence of an important length scale, which may be considered as
the mean shock wave separation. The mute point is whether this is the
single dominant length in the problem.  If so, then one has dynamical
scaling in its simplest form, and many quantities may be subsequently
obtained by scaling arguments. What is lacking in the previous work on
BE is an explicit solvable case in which scaling is seen to emerge
cleanly. In order to achieve this it is necessary to calculate the
form of some correlation function, which entails more difficulties
than studying for instance the mean energy decay. Our intention here
is to present such a calculation for a class of initial conditions in
which the velocity potential is a bounded, discontinuous, random
function.  In this case exact calculations are possible and we may
extract the form of the velocity-velocity correlation function $E({\bf
  r},t)$ for arbitrary spatial dimension $d$. We find that there
exists dynamical scaling, but that it is controlled by two length
scales, rather than one -- a diffusive scale $l_{D}$ for large
distances, and a super-diffusive scale $L(t)$ for small distances.
The details underlying this remark will be given below, but the
important conceptual point is that if two length scales are playing a
scaling role, then their ratio (which is, of course, dimensionless)
may play a hidden role in subsequent scaling arguments. Thus, simple
dimensional analysis is likely to fail. We shall see an explicit
demonstration of this as we proceed.

The outline of the paper is as follows. In the next section we
introduce BE, discuss various choices of initial conditions, and
briefly describe a few analytic steps which are required before the
calculation proper.  By adopting the initial condition mentioned
above, along with some new analytic methods, we are able to calculate
$E({\bf r},t)$, and we give explicit forms for its asymptotic
behaviour for small, and large distances. This is presented in the
lengthy section IV, section III being a warm-up exercise to calculate
the mean energy decay (which has previously appeared in
print\cite{esnew}). In section V we make a (formally exact) connection
between BE and a popular `toy model' in disordered systems theory. We
then present simple scaling arguments for the toy model which yield
partial agreement with the more complicated scaling picture that
emerges from sections III and IV. The paper ends with our conclusions.

\section{Definition of the model}

BE is a partial differential equation written in terms of a velocity
field ${\bf v}({\bf x},t)$:
\begin{equation}
\label{bev}
\partial _{t} {\bf v} = \nu \nabla ^{2} {\bf v} - 
({\bf v}\cdot \nabla){\bf v} \ .
\end{equation}
The field is taken to be irrotational, which allows one to express the
equation solely in terms of the velocity potential $\phi $ defined via
${\bf v} = -\nabla \phi$.  Explicitly one has
\begin{equation}
\label{bep}
\partial _{t} \phi = \nu \nabla ^{2}\phi + 
(1/2) (\nabla \phi)^{2} \ .
\end{equation}
The equation is most commonly discussed in one spatial dimension, in
the spirit of its application to non-linear waves\cite{gurb}.
However, the $d$-dimensional generalization given above is the
canonical choice.  One is interested in the evolution of the velocity
field from some given initial condition, in the limit of vanishing
viscosity, i.e. $\nu \rightarrow 0$. This leads to strongly non-linear
behaviour, otherwise known as the strong turbulence limit. We shall
make this limit more explicit in terms of a dynamic Reynolds number as
we proceed.

The initial conditions we shall study are random functions ${\bf
  v}_{0}({\bf x})$, and as such are defined in terms of a distribution
function $P[{\bf v}_{0}]$. Naturally, there is an enormous choice
available for $P$. Burgers\cite{b2} studied perhaps the most natural,
namely a Gaussian distribution of velocities with $\delta $-function
correlations:
\begin{equation}
\label{burgersic}
P_{B}[{\bf v}_{0}] \sim \exp \left ( -(1/2D)\int d^{d}x \ {\bf
  v_{0}}^{2} \right ) \ .
\end{equation}
His analysis was confined to $d=1$ where a controlled analytic study
was possible. The main result to emerge was that the velocity field
forms into shock waves separated by smooth regions, and that the
shocks become more dilute as time proceeds; the mean shock wave
separation increasing as $L_{s} \sim t^{2/3}$.  Dimensional arguments
indicate that above $d=2$, the asymptotic properties are dominated by
diffusion (i.e. the shock waves disappear and the field diffusively
vanishes), so that the dominant length scale is then a diffusion scale
growing as $t^{1/2}$. In precisely two dimensions\cite{es2}, diffusion
is still the dominant process, but logarithmic corrections are
expected for quantities such as the mean energy decay ${\cal E}(t)
\equiv \langle {\bf v}^{2} \rangle$ (where here and henceforth, angled
brackets indicate an average over the ensemble of initial conditions).
Generalizations of the Gaussian form of the initial conditions (for
example, defining different power spectra in Fourier space) have been
studied previously\cite{gurb,kida,gurb1,sinai}, and scaling arguments
have provided a broad classification for the time-dependence of the
length scale $L_{s}(t)$

One may also consider initial distributions in terms of the velocity
potential. A particular class of these is to divide the system into
cells of size $l^{d}$, and to assign a value of $\phi _{0}$
independently within each cell\cite{esnew,es2}.  In this case we may
write the distribution as
\begin{equation}
\label{disc}
P[{\bf v}_{0}] = \prod \limits _{cells} p(\phi_{0,cell}) \ .
\end{equation}
It is important to distinguish between cell distributions $p$ which
allow bounded or unbounded values of their argument. We shall see that
distributions of the former class (such as a top-hat function, or a
cut-off exponential distribution) constitute a particular universality
class, whereas those of the latter (such as a Gaussian or a power-law
distribution) have different scaling properties.  In the present work
we shall be interested solely in random initial conditions of the
former type, by demanding the cell distribution function to be defined
only for a finite range of the velocity potential. Furthermore, one
may show that all such distributions lead to the same asymptotic
behaviour when the width of the distribution is large (but still
finite), and we therefore concentrate on the simplest case, namely a
top-hat function. [This is strictly true only for distributions
which fall to zero discontinuously. Distributions which vanish 
at some finite value of their argument in a smooth manner require
a separate analysis.] Explicitly we choose
\begin{equation}
\label{tophat}
p(\phi_{0}) = {\theta (\Phi-|\phi_{0}|) \over 2\Phi} \ ,
\end{equation}
where $\theta (z)$ is the Heaviside unit function\cite{jeff}.

Analytic progress has been possible in BE over the years, since for a
given initial condition, one may exactly integrate the equation. This
is due to the Hopf-Cole\cite{hopf,cole} transformation which
linearizes BE. By defining $w({\bf x},t) = \exp [\phi ({\bf
  x},t)/2\nu]$ and substituting into eq.(\ref{bep}), one may see that
$w$ satisfies the linear diffusion equation which is immediately
solved in terms of the heat kernel $g({\bf x},t) = (4\pi \nu
t)^{-d/2}\exp[-x^{2}/4\nu t]$.  Re-expressing the solution in terms of
the velocity potential, one has the explicit solution of BE in the
form
\begin{equation}
\label{exsol}
\phi ({\bf x},t) = 2\nu \ln \left \lbrace \int d^{d}y \ g({\bf x}-{\bf
  y},t) \exp \left [ \phi_{0}({\bf y})/2\nu \right ] \right \rbrace \ 
.
\end{equation}

The main analytic effort is to now perform averages over the initial
distribution $P$. We shall accomplish this by making the following
integral representation of the logarithm function in the above
expression:
\begin{equation}
\label{intrepresen}
\ln (z) = \int \limits _{0}^{\infty} {du\over u}(e^{-u} - e^{-uz}) \ .
\end{equation}
(This representation has proven useful\cite{derr} in calculations in
disordered systems theory as an alternative to the replica method, and
has also been used previously in problems related to
BE\cite{esnew,es2,new}).  We therefore have the solution of BE in the
form
\begin{equation}
\label{exsol1}
\phi ({\bf x},t) = 2\nu \int \limits _{0}^{\infty} {du\over u}[e^{-u}
- \psi (u,{\bf x},t)] \ ,
\end{equation}
where
\begin{equation}
\label{psif}
\psi = \exp \left \lbrace -u \int d^{d}y \ g({\bf x}-{\bf y},t) \exp
\left [ \phi_{0}({\bf y})/2\nu \right ]\right \rbrace \ .
\end{equation}

Our main focus in this work is to calculate the velocity-velocity
correlation function defined by
\begin{equation}
\label{energycorrfn}
E({\bf r},t) = (1/2)\langle {\bf v}({\bf r},t) \cdot {\bf v}({\bf
  0},t) \rangle
\end{equation}
which may be easily related (with the aid of translational invariance)
to the two-point correlation function for the velocity potential via
\begin{equation}
\label{corr}
E({\bf r},t) = (1/4) \nabla ^{2} C({\bf r},t) \ ,
\end{equation}
where
\begin{equation}
\label{corr1}
C({\bf r},t) = \langle [\phi({\bf r},t) - \phi({\bf 0},t)]^{2} \rangle
\ .
\end{equation}

The mean energy decay is given simply by ${\cal E}(t) \equiv E({\bf
  0},t)$, and we shall present a condensed derivation of this quantity
in the next section, before tackling the much harder task of
calculating $C({\bf r},t)$. Results for $E({\bf 0},t)$ have been
presented before\cite{esnew}, but it is useful to sketch the
derivation here in order to set up the necessary formalism required in
section IV, along with revealing the important time scale in the
problem.

\section{Calculation of the energy decay}

In previous studies of BE\cite{gurb,b2,kida}, it is more common to
infer the energy decay from a scaling argument, once one has
calculated the important dynamic length scale $L_{s}(t)$. On
dimensional grounds one would like to infer ${\cal E} \sim
L_{s}^{2}/t^{2}$. This scaling relation certainly holds true in many
situations, but it is by no means a universal result. We shall attack
the problem from the opposite direction, by first calculating the
energy decay explicitly. We shall then read off important length
scales from the correlation function in section IV -- comparing the
two independent results will then allow us to see if dimensional
analysis holds for our particular choice of the initial distribution.

Expressing ${\cal E}$ in terms of the velocity potential, one may see
from averaging eq.(\ref{bep}) over the initial distribution, that
\begin{equation}
\label{energyq}
{\cal E} = \partial _{t} \langle \phi ({\bf x},t) \rangle \ .
\end{equation} 
So in order to determine the energy decay, we need only calculate the
mean velocity potential, which in turn is related to the average of
the function $\psi $ from eq.(\ref{exsol1}). In fact a very similar
function will be central in the evaluation of the correlation
function, so it is useful to dedicate a few lines to deriving an
explicit expression for $\langle \psi \rangle$.

In order to perform the average it is necessary to impose a lattice
scale $a$ -- this is because generally the initial condition average
has the form of a functional integral, which is only strictly defined
on a lattice. We shall find that for all but the shortest times (set
by $t_{0}$ -- the time for diffusion over the cell size l) this scale
$a$ disappears from all physical quantities, and is replaced by the
cell scale $l$ which defines the correlation scale of the initial
conditions. Explicitly we define a diffusion length $l_{D} = (4\nu
t)^{1/2}$ and work in the limit $a \le l \ll l_{D}$.  In other words,
the spatial smearing of the heat kernel is much greater than the cell
size. Performing the initial condition average over $\psi $ using the
distribution defined by eqs.(\ref{disc}) and (\ref{tophat}), we find
\begin{equation}
\label{psiav1}
\ln \langle \psi \rangle = l^{-d} \int d^{d}y \ln \left \lbrace
{1\over K_{0}} \left [ E_{1}\left ( ul^{d}g({\bf
  y},t)e^{-K_{0}/2}\right ) - E_{1}\left ( ul^{d}g({\bf
  y},t)e^{K_{0}/2}\right ) \right ] \right \rbrace \ ,
\end{equation}
where $E_{1}(z)$ is the exponential integral\cite{as} and we have
defined $K_{0}$ to be the effective Reynolds number at zero time: i.e.
$K_{0} = ({\rm typical \ velocity} \times {\rm typical \ length})/\nu
= \Phi/\nu$.  (We refer the reader to Appendix A where the initial
condition average is performed explicitly.)  We may simplify this
expression in two steps. Firstly we make the rescaling $u \rightarrow
u(\pi^{1/2} l_{D}/l)^{d}e^{-K_{0}/2}$ and change the integration
variable to $s=y^{2}/4\nu t$. Secondly, we impose the strong
turbulence limit by taking $K_{0} \gg 1$. This leads us to
\begin{equation}
\label{psiav2}
\ln \langle \psi \rangle = -{(\pi ^{1/2}l_{D}/l)^{d} \over K_{0}\Gamma
  (d/2+1)} \ L_{d/2}(u) \ + O(1/K_{0}^{2}) \ ,
\end{equation}   
where $\Gamma (z)$ is the gamma function\cite{as}, and
\begin{equation}
\label{intdef}
L_{p}(u) \equiv \int \limits _{0}^{\infty} ds \ s^{p}
[1-\exp(-ue^{-s})]
\end{equation}
We refer the reader to Appendix B where it is shown that the integral
may be evaluated for both small and large values of $u$ with the
result
\begin{equation}
\label{integral}
\nonumber L_{p}(u) = \left \{
\begin{array}{ll}
  u\Gamma (p+1) \ [1 - 2^{-(p+2)}u + O(u^{2})] \ , & u \ll 1 \\ {[\ln
    (u)]^{p+1}\over (p+1)} + \gamma [\ln (u)]^{p} + O([\ln (u)]^{p-1})
  \ , & u \gg 1 \ .
\end{array}
\right.
\end{equation}

In order to find the mean velocity potential we must perform the
$u$-integral as given in eq.(\ref{exsol1}). One may see that the
$u$-integral is dominated by $u \ll 1 \ (\gg 1)$ when the ratio
$(l_{D}/l)^{d}/K_{0} \gg 1 \ (\ll 1)$ The former case occurs for very
large times, and on performing the $u$-integral one obtains a
diffusion result -- i.e. ${\cal E} \sim t^{-(d/2+1)}$. So, there
exists a cross-over time $t_{c} \sim (l^{2}/\nu)K_{0}^{2/d}$ beyond
which the non-linearity is irrelevant and the velocity potential
relaxes according to diffusion. By taking the initial Reynolds number
to be arbitrarily large, we may push $t_{c}$ to arbitrarily late
times. The interesting non-linear regime occurs for $t_{0} \ll t \ll
t_{c}$ in which case one must perform the $u$-integral using the
large-$u$ asymptotic form for $\langle \psi \rangle $. In this case
one finds (using the variable change $\sigma = [\ln (u)]^{d/2+1}$, and
imposing a lower cut-off of $O(1)$ to the $u$-integral)
\begin{equation}
\label{energy}
{\cal E} = C_{d} \ {(K_{0}l^{d}l_{D}^{2})^{2/(d+2)} \over t^{2}} \sim
t^{-\sigma} \ ,
\end{equation}
where $\sigma = 2(d+1)/(d+2)$ and $C_{d}$ is a complicated
$d$-dependent constant.

We may reinterpret this expression by defining a time-dependent
Reynolds number $K(t)$. For a typical velocity we take the square root
of the mean energy decay, and for a typical (large) length scale we
take $l_{D}$ (which we will justify {\it a posteriori} in section IV).
Then we have
\begin{equation}
\label{reynolds}
K(t) = \left [ C'_{d} K_{0} \left ({l\over l_{D}}\right )^{d} \right ]
^{1/(d+2)} \ ,
\end{equation}
where the constant $C'_{d} = \pi^{-d/2} \Gamma(d/2+2)$ has been chosen
for future convenience.  We can see that $K(t)$ decays from its
initial (very) large value with the power law $t^{-d/2(d+2)}$ until it
becomes of order unity when $t \sim t_{c}$.  In this non-linear regime
(defined by $K(t) \gg 1$) we may write the energy decay in the form
\begin{equation}
\label{rew}
{\cal E} \sim {l_{D}^{2} \over t^{2}} \ K(t)^{2} \ ,
\end{equation}
which is cast into the form `expected' from dimensional analysis,
except that the dimensionless (but time dependent) Reynolds number is
also present, which invalidates the prediction for the time dependence
of ${\cal E}$ from dimensional considerations alone.

The introduction of the time-dependent Reynolds number is useful, but
must be justified by independently proving that $l_{D}$ is the typical
(large) length scale in the non-linear regime. Alternatively one could
insist on the dimensional prediction, in which case one would infer
the important length scale to be $L_{s} \sim l_{D}K(t)$. (In fact, we
shall see that the dynamic length scale $\sim l_{D}/K(t)$.) To place
these results and speculations into a proper context one is forced to
evaluate the scaling properties of the correlation function, which is
a much more difficult task than the calculation of the mean energy
decay.

\section{Calculation of the correlation function}

This section constitutes the heart of the paper in that we present the
exact asymptotic forms for the correlation function $C({\bf r},t)$ in
the non-linear regime $t_{0} \ll t \ll t_{c}$. Unfortunately, in order
to arrive at the required result, one must wade through a very long
and technical calculation. So as not to burden the reader with
details, all technical remarks will be relegated to appendices, with
only the general flow of the analysis described in the main text.
Firstly, we shall derive a general expression for $C$ in the
non-linear regime. In the subsequent sub-sections, we shall then
analyse the asymptotic properties of $C$ in the limits of $r \ll
l_{D}$ and $r \gg l_{D}$.  As hinted at before, the main result of
this analysis is the emergence of a new length scale which describes
the small distance behaviour of the correlation function.

It is convenient to define $C$ in a symmetric way (cf.
eq.(\ref{corr1})):
\begin{equation}
\label{symmcorr}
C({\bf r},t) = \langle [\phi(-{\bf r}/2,t) - \phi({\bf r}/2,t)]^{2}
\rangle = 2\langle \phi({\bf 0},t)^{2} \rangle - 2\langle \phi(-{\bf
  r}/2,t)\phi({\bf r}/2,t)\rangle \ .
\end{equation}
By utilizing the integral representation of the logarithm function
twice, we may rewrite the bilinear combinations of velocity potentials
in terms of integrals.  This yields
\begin{equation}
\label{corrfn}
C({\bf r},t) = 8\nu^{2} \int \limits _{0}^{\infty} {du \over u} \int
\limits _{0}^{\infty} {dv \over v} \ \left [ \Psi (u,v,{\bf 0},t) -
\Psi (u,v,{\bf r},t) \right ] \ ,
\end{equation}
where
\begin{equation}
\label{corrfn1}
\Psi (u,v,{\bf r},t) = \left \langle \exp \left [- \int d^{d}y \ \left
( ug({\bf y}-{\bf r}/2,t) + vg({\bf y}+{\bf r}/2,t) \right )
e^{\phi_{0}({\bf y})/2\nu} \right ] \right \rangle \ ,
\end{equation}
and we have utilized the property of translational invariance.

In an analogous fashion to the averaging performed in the previous
section, the average over the initial conditions may be performed in a
straight-forward manner (see Appendix A), yielding a complicated
expression for $\Psi$ in terms of the exponential integral function.
However, great simplification may be made by taking the limit $K_{0}
\gg 1$. In this case we reduce $\Psi $ to the form
\begin{equation}
\label{psifn}
\ln [ \Psi (u,v,{\bf r},t) ] = -\epsilon I(u,v,{\bf R}) \ + \ 
O(1/K_{0}^{2}) \ ,
\end{equation}
where $\epsilon = \Gamma (d/2+2)/K(t)^{d+2} \ll 1$, and
\begin{eqnarray}
\label{ifn}
\nonumber I(u,v,{\bf R}) = {2\over d\pi^{d/2}}\int d^{d}y & \ & \Biggl
\lbrace y^{2} - {{\bf y} \cdot {\bf R}\over 2} \left [ {ue^{-({\bf
      y}-{\bf R}/2)^{2}} - ve^{-({\bf y}+{\bf R}/2)^{2}} \over
  ue^{-({\bf y}-{\bf R}/2)^{2}} + ve^{-({\bf y}+{\bf R}/2)^{2}} }
\right ] \Biggr \rbrace
\\ \times & \ & \Biggl \lbrace 1 - \exp \left [ -ue^{-({\bf y}-{\bf
    R}/2)^{2}} -ve^{-({\bf y}+{\bf R}/2)^{2}} \right ] \Biggr \rbrace
\ ,
\end{eqnarray}
and we have defined ${\bf R} \equiv {\bf r}/l_{D}$.

At this point of the discussion it is convenient to consider the small
and large distance behaviours of $C$ separately.

\subsection{Small distance scaling}

To ascertain the small distance properties of $C$ we need to
perturbatively evaluate the above integrals in a power series in $R
\ll 1$. Although one may attempt this directly on the form of the
integrals as given by eqs.(\ref{corrfn1}) and (\ref{ifn}), it is far
more efficient to transform the function $I$ beforehand into a natural
power series in $R^{2}$. The procedure for this is described in
Appendix C, with the result
\begin{equation}
\label{infsum}
I(u,v,{\bf R}) = \sum \limits _{p=0}^{\infty} (R^{2})^{p} \ F_{p}(u,v)
\ ,
\end{equation}
where the functions $F_{p}$ have the integral form
\begin{equation}
\label{funcfp}
F_{p}(u,v) = {(-uv\partial _{u}\partial _{v})^{p} \over \Gamma (p+1)
  \Gamma (p+d/2+1)} \ L_{p+d/2}(u+v) \ .
\end{equation}

We are interested in the non-linear regime $K(t) \gg 1$, and in this
case the $(u,v)$ integrals are dominated by $u \gg 1$ and $v \gg 1$.
Therefore we expand the integral $L_{p+d/2}$ appearing in
eq.(\ref{funcfp}) in powers of $\Delta \equiv \ln (u+v) \gg 1$ (see
Appendix B).  The function $F_{p}$ may now be expressed as
\begin{equation}
\label{aaah}
F_{p}(u,v) = {(-1)^{p} \over \Gamma (p+1) \Gamma (p+d/2+1)} \ \left [
{\chi_{p}(u,v;p+d/2+1) \over (p+d/2+1)} + \gamma \chi_{p}(u,v;p+d/2) +
O(\Delta ^{p+d/2-2}) \right ] \ ,
\end{equation}
where we have defined
\begin{eqnarray}
\label{aaah1}
\nonumber \chi_{p}(u,v;q) & = & (uv \partial_u \partial_v)^{p} \ 
\Delta ^{q} \\ & = & f_{p}(u,v;q)\Delta ^{q-1} + g_{p}(u,v;q) \Delta
^{q-2} + O(\Delta ^{q-3}) \ .
\end{eqnarray}
More details of these steps, along with the explicit form of the
coefficients $\lbrace f_{p} \rbrace$ and $\lbrace g_{p} \rbrace$ may
be found in Appendix D.

Now that we have a workable series for $I$ in the non-linear regime,
it is possible to expand $\Psi (u,v,{\bf r},t)$ in powers of $R^{2}$,
such that the coefficients are various combinations of the functions
$F_{p}$.  The integrals over $u$ and $v$ may then be performed (see
Appendix E for details), and one has the final result
\begin{eqnarray}
\label{final}
\nonumber {C({\bf r},t)\over 8\nu^{2} } & = & {2 \over (d+2) } \Gamma
\left ( {d+4 \over d+2} \right ) K(t)^{2} \ R^{2} \\ \nonumber
\\
\nonumber & & - \ \Biggl \lbrace \ R^{2} + {(d+3)\over 3(d+2)^{2}}
\Gamma \left ( {d+4\over d+2} \right ) K(t)^{2} \ R^{4} - {4 (d+5)
  \over 45(d+2)^{2}(d+4)} \Gamma \left ( {d+6 \over d+2} \right )
K(t)^{4} \ R^{6} \\ \nonumber & & \ \ \ \ \ \ \ + {4 (d+5)(d+7) \over
  63(d+2)^{2}(d+4)^{2}(d+6)} \Gamma \left ( {d+8 \over d+2} \right )
K(t)^{6} \ R^{8} + O(R^{10}) \ \Biggr \rbrace \\ \nonumber
\\
& & + \ \cdots
\end{eqnarray}

Several remarks are now in order. Firstly, the above result is given
(after much effort) to quite high order in $R^{2}$. One is obliged to
do this to unambiguously determine the scaling properties of the
correlation function. Secondly, the result for $C$ has been written in
such a way as to stress the form of the scaling. It turns out that the
dominant term at each order of $R^{2}$ vanishes exactly, except for
the dominant term {\it at} order $R^{2}$ -- this explains why this
term stands alone in the above expression. The sub-dominant terms from
each order are non-zero, and are grouped together within the braces.
The remaining terms play no role in determining the leading scaling
behaviour and are indicated by the ellipsis.  The fact that the
dominant terms vanish means that the leading $R^{2}$ term can play no
part in the scaling form of the correlation function. However, the
terms in braces have a natural scaling form which allows us to read
off a dynamic length scale. Explicitly we may recast the above
expression into the scaling form (neglecting constants)
\begin{equation}
\label{scaling}
C({\bf r},t) \sim \nu^{2}\left [ \left ( {{\bf r}\over L(t)} \right
)^{2} \ + \ \left ( {{\bf r} \over l_{D}}\right ) ^{2} S\left ( {{\bf
    r}\over L(t)} \right ) \right ] \ ,
\end{equation}
where $S(x)$ is the scaling function, and the dynamic length scale is
$L(t) = l_{D}/K(t) \sim t^{\alpha }$ with $\alpha = (d+1)/(d+2)$.

We see that in the non-linear regime, the dynamic length scale is much
smaller than the diffusive scale $l_{D}$, although it is growing
faster. This gives us another view of the cross-over from non-linear
to linear evolution; i.e. the dynamic Reynolds number becomes of order
unity when the scale $L(t)$ becomes of the same order as $l_{D}$.

As a final remark in this section, we may obtain the velocity-velocity
correlation function from eq.(\ref{scaling}) with the use of
eq.(\ref{corr}).  One finds
\begin{equation}
\label{velscaling}
E({\bf r},t) \ \sim \ {\cal E}(t) \ + \ \left ({\nu \over l_{D}}\right
)^{2} {\tilde S} \left ( {{\bf r}\over L(t)} \right ) \ .
\end{equation} 
Again, it is interesting to see that the mean energy decay ${\cal
  E}(t)$ is not part of the scaling function, which explains the
difficulties encountered in the previous section with simple
dimensional analysis. It remains to show the scaling importance of the
diffusive scale - this will be accomplished in the next section.

\subsection{Large distance scaling}

The scaling form for the correlation function for very large $|{\bf
  r}|$ may be obtained with relatively little effort. Starting with
$C({\bf r},t)$ expressed in terms of the function $I(u,v,{\bf R})$ as
given in eqs.(\ref{corrfn}), (\ref{psifn}) and (\ref{ifn}), we may
express $I$ by the following series (cf. Appendix B):
\begin{equation}
\label{larger}
I(u,v,{\bf R}) = \sum \limits _{n=1}^{\infty} {(-1)^{n+1}\over n!} \ 
n^{-(d/2+1)} \ \sum \limits _{m=0}^{n} C^{n}_{m} u^{m}v^{n-m} \ \exp
\left [ -{m \over n} \ (n-m) R^{2} \right ] \ .
\end{equation} 
As $C({\bf r},t)$ is non-zero for $|{\bf r}| \rightarrow \infty$, it
is convenient to measure correlations with respect to the asymptotic
value $C(\infty,t)$. Thus we define
\begin{eqnarray}
\label{dc}
\nonumber \delta C \equiv C(\infty,t)-C({\bf r},t) & = & 8\nu^{2}\int
\limits _{0}^{\infty} {du \over u} \int \limits _{0}^{\infty} {dv
  \over v} \bigl [ \Psi (u,v,{\bf r},t)-\Psi(u,v,\infty,t) \bigr ] \\ 
& = & 8\nu^{2}\int \limits _{0}^{\infty} {du \over u} \int \limits
_{0}^{\infty} {dv \over v} \Bigl \lbrace \exp [-\epsilon I(u,v,{\bf
  R})] - \exp [-\epsilon I(u,v,\infty) ] \Bigr \rbrace \ .
\end{eqnarray}

From eq.(\ref{larger}) it is easy to see
\begin{equation}
\label{iinf}
I(u,v,\infty) = \sum \limits _{n=1}^{\infty}{(-1)^{n+1}\over n!} \ 
n^{-(d/2+1)} \ (u^{n} + v^{n}) = {[L_{d/2}(u)+L_{d/2}(v)] \over \Gamma
  (d/2+1)} \ ,
\end{equation}
where we have rewritten the sum in terms of the familiar integral
$L_{d/2}$ (using the integral representation shown in
eq.(\ref{intrepres2})).  In the non-linear regime, we are interested
in the large $(u,v)$ behaviour, which according to eq.(\ref{integral})
gives us
\begin{equation}
\label{iinf1}
I(u,v,\infty) \sim {1 \over \Gamma (d/2+2)} \left \lbrace [\ln
(u)]^{d/2+1} + [\ln (v)]^{d/2+1} \right \rbrace \ .
\end{equation}

Returning to eq.(\ref{larger}) the large-$|{\bf R}|$ form for $I$ may
be written as
\begin{equation}
\label{correct}
I(u,v,{\bf R}) = I(u,v,\infty) - 2^{-(d/2+1)} \ uve^{-R^{2}/2} +
\cdots \ .
\end{equation}
Substituting this result into eq.(\ref{dc}) gives the leading term of
$\delta C$ as
\begin{equation}
\label{leading}
\delta C \sim {4\nu^{2}\epsilon \over 2^{d/2}} \ J(K)^{2} \ 
e^{-R^{2}/2} \ ,
\end{equation}
where
\begin{equation}
\label{integ}
J(K) = \int \limits _{c}^{\infty} du \ \exp \left \lbrace
-K(t)^{-(d+2)} [\ln (u)]^{d/2+1} \right \rbrace \ ,
\end{equation}
($c$ being a number of order unity).

The integral may be performed by steepest descents in the non-linear
regime (details in Appendix B) with the result that $\delta C$ has the
final asymptotic form (neglecting overall constants)
\begin{equation}
\label{finalform1}
\delta C \sim \nu ^{2} K(t)^{(4-d^{2})/d} \exp[C_{d}''K(t)^{2(d+2)/d}]
\ \exp[-r^{2}/2l_{D}^{2}] \ ,
\end{equation}
where $C_{d}'' = d(d/2+1)^{-(1+2/d)}$.  We see from this expression
that the diffusion scale $l_{D}$ is the natural scaling length for the
correlation function, for very large distances. One may ascertain the
range of validity of the above expression by calculating the
contribution from the next term in the series (from
eq.(\ref{larger})), and one finds that the above form is valid for
$|{\bf r}| \gg lK_{0}^{1/d} \ ( = (\nu t_{c})^{1/2} \gg l_{D})$.

Before ending this rather technical section, we shall recap the main
results obtained.  By performing an exact analysis on the correlation
function $C({\bf r},t)$ in the non-linear regime, we have been able to
confirm that there indeed exists dynamical scaling, albeit of a rather
subtle type. The small distance properties of $C$ are governed by a
scaling length $L(t) \sim l_{D}/K(t)$, but the dominant term in $C$ is
singular -- i.e. it may not be included into the scaling form. This
indicates why the form of the mean energy decay found in the previous
section was not obtainable by simple dimensional analysis. The scale
$L(t)$ is much smaller than the diffusion scale, but grows faster --
the non-linear regime crosses over to simple diffusion when these two
length scales become compatible. The large distance scaling was found
to be more conventional, in that the dominant part of $C$ (with
respect to its asymptotic value) is a simple function of $r/l_{D}$,
albeit with a complicated prefactor; thus indicating that $l_{D}$ acts
as the dynamic scale for the correlation function at very large
distances.

\section{Scaling arguments}

This section has two purposes. Firstly, we shall show an exact mapping
between the solution of BE, and the free energy of a popular toy
model\cite{tm} in the field of disordered systems. Secondly, we shall
perform some simple scaling calculations\cite{bray1} on the latter
model to extract the form of the mean energy decay ${\cal E}(t)$ in
the original BE problem.  These scaling calculations are very similar
in spirit to the original calculations of Burgers\cite{b2} and
Kida\cite{kida}.

The toy model in question is simply described. Consider a single
particle in a potential composed of a harmonic background plus a
random potential $V({\bf x})$. If the particle is in contact with a
thermal reservoir, we may write the partition function for the
particle as
\begin{equation}
\label{partfn}
Z = \left ( {\beta \mu \over 2\pi} \right )^{d/2} \int d^{d}x \ \exp
\left \lbrace -\beta \left [ {\mu \over 2} x^{2} +V({\bf x}) \right ]
\right \rbrace \ ,
\end{equation}
where $\beta$ is the inverse temperature, and we have normalized $Z$
with respect to the harmonic background.  For a given realization of
the disorder potential $V$, we may calculate the free energy of the
particle from $F_{V} = -(1/\beta) \ln (Z)$.

At this stage we compare the expression for the free energy, with the
exact solution of BE (evaluated at the origin) as given by
eq.(\ref{exsol}).  We see that there exists an exact equivalence
between the two quantities, if one makes the following connections:
$\phi({\bf 0},t) \leftrightarrow -F_{V}, \ 2\nu \leftrightarrow
1/\beta, \ t \leftrightarrow 1/\mu, \ {\rm and} \ \phi _{0}({\bf x})
\leftrightarrow -V({\bf x})$. This correspondence holds regardless of
the particular distribution of initial conditions (or equivalently,
disorder).

To proceed with the description of the toy model, two quantities one
may be interested in calculating are the quenched free energy $F =
\langle F_{V} \rangle$, and the mean square displacement of the
particle
\begin{eqnarray}
\label{msdisp}
\nonumber \langle x^{2} \rangle & = & \left \langle {1 \over Z} \int
d^{d}x \ x^{2} \exp \left \lbrace -\beta \left [ {\mu \over 2} x^{2}
+V({\bf x}) \right ] \right \rbrace \right \rangle\\ & = & \left
\langle -(2/\beta) \partial _{\mu} \ \ln (Z) \right \rangle \ = \ 
2\partial _{\mu} F \ .
\end{eqnarray}
By utilizing the correspondence with BE, we may relate the mean square
displacement to the quantity ${\cal E}(t)$ in BE. Explicitly we write
\begin{eqnarray}
\label{corresp}
\nonumber \langle x^{2} \rangle & = & 2\partial _{\mu} F = -2\partial
_{1/t}\langle \phi ({\bf 0},t) \rangle \\ & = & 2t^{2}\partial
_{t}\langle \phi ({\bf 0},t) \rangle = 2t^{2}{\cal E}(t) \ ,
\end{eqnarray}
where we have made use of eq.(\ref{energyq}).  So we have been able to
show that the dimensional prediction for the mean energy decay, namely
${\cal E}(t) \sim L_{s}(t)^{2}/t^{2}$, has a formal interpretation in
terms of the toy model, so long as we interpret $L_{s}(t)$ as the
root-mean-square displacement of the particle.

We shall now derive an approximate form for $L_{s}(t)$ within the toy
model formulation.  Consider first, the top-hat distribution that has
been the subject of the present work.  We take $P[V] = \prod \limits
_{cells} p(V_{cell})$, with $p(V)=\theta(D-|V|)/2D$.  The strong
turbulence limit of BE corresponds to the zero temperature limit of
the toy model. In this case, the particle will be trapped in the
lowest potential energy minimum, within a given realization. In this
case we may estimate the excursion of the particle by calculating the
probability $q(r,U^{*})$ for the lowest potential site to be located
at a distance $r$ from the origin, and to have an energy $U^{*}$. This
will be proportional to the probability that all sites within a radius
$r$ of the origin have a potential energy higher than $U^{*}$.

For a general potential distribution $p(V)$, we may write
\begin{equation}
\label{potene}
q(r,U^{*}) \sim p(U^{*}-\mu r^{2}/2) \ \prod \limits _{|{\bf x}|<r} \ \ 
\int \limits _{U^{*}-\mu x^{2}/2}^{\infty} dV_{\bf x} \ p(V_{\bf x}) \ 
.
\end{equation}
In the BE analogy we are interested in long times, so within the toy
model we need to take $\mu $ to be small, i.e., the harmonic
background is taken to be very `flat'.
 
Restricting our attention to the top hat distribution, we see that a
flat harmonic background implies that the minimal energy $U^{*}$ will
be close to the lower bound of the random potential $-D$. We therefore
set $U^{*}=-D+\delta U$, where $|\delta U| \ll D$. The above
expression then reduces to
\begin{equation}
\label{potene1}
q(r,U^{*}) \sim {1\over 2D} \ \prod \limits _{|{\bf x}|<r} \left [1 -
{\delta U \over 2D} + {\mu x^{2} \over 4D} \right ] \ .
\end{equation}
It is now straightforward to exponentiate the term in square brackets,
so as to transform the product over ${\bf x}$ as a spatial integral in
the exponential. Taking $\mu r^{2} \ll D$ (which is justified {\it a
  posteriori} by the form of $r_{\rm{typ}}$ given below) allows the
integral to be performed simply, yielding the final result
\begin{equation}
\label{distfin}
q(r,\delta) \sim {1 \over 2D} \exp \left [ -c_{1}(d){\delta \over D}
\left ( r \over l \right )^{d} + c_{2}(d) { \mu r^{d+2} \over l^{d} D
    } \right ] \ ,
\end{equation}
where $c_{1}$ and $c_{2}$ are constants.  For this distribution of
potential minima at distance $r$ from the origin, we can read off a
scaling relation between the typical value of $r$ and $\mu $; namely
\begin{equation}
\label{scalingrel}
r_{\rm typ} \sim \left ( {l^{d}D \over \mu} \right ) ^{1/(d+2)} \ .
\end{equation}
Making the correspondence with BE, we identify $r_{\rm typ}
\leftrightarrow L_{s}(t)$ and $\mu/D \leftrightarrow
(l_{D}^{2}K_{0})^{-1}$. Combining the above result with
eq.(\ref{corresp}), we see that we have derived the correct form of
the mean energy decay as calculated previously in section III, cf.
eq.(\ref{energy}); although the precise value of the prefactor may not
be obtained by this simple scaling argument.

The above result may also be cast into the form $L_{s}(t) \sim
l_{D}K(t)$. This is guaranteed under the scaling hypothesis, but what
is interesting is that such a length scale plays no role in the actual
dynamical scaling as defined by the behaviour of the two-point
correlation function. In other words, although we may calculate
$L_{s}$ (or rather $r_{\rm typ}$) from scaling considerations of the
toy model, this length scale is not a dynamic scaling length -- for
instance, it could not be used to collapse the correlation function in
a scaling plot.

To end this section, we mention that the toy model may be analysed for
other types of distribution. If one takes the disorder distribution to
be of the cellular type, with $p$ a gaussian, then one may rederive
the result of Kida\cite{kida}; namely that $L_{s}(t)$ is a diffusive
scale up to logarithmic corrections.  Alternatively one may consider
the toy model in $d=1$ with a disorder distribution corresponding to
the original Burgers choice, namely $P[V] \sim \exp [ -\int dx \ 
(dV/dx)^{2}]$.  This particular scenario has been studied in detail
recently, using a new replica approach\cite{mez}. Although the
essential Burgers scaling ($L_{s}(t) \sim t^{2/3}$) is easily
recovered, the calculation of prefactors is more difficult. It is an
important test of various approaches as to whether they can
quantitatively predict the correct prefactor.  As far as we are aware,
this is still an open problem, although there are a number of
approximate values in the literature\cite{b2,esnew,mez}.

\section{Conclusions}

This paper has been concerned with proving dynamical scaling for
Burgers equation with random initial conditions. Exact calculations
have been possible for a distribution of the initial velocity
potential which has a large but finite region of support, and is
uncorrelated in space. In section III we calculated the mean energy
decay ${\cal E}(t)$ in the non-linear regime (i.e. in the temporal
regime in which the Reynolds number $K(t) \gg 1$). Explicitly we found
\begin{equation}
\label{rew1}
{\cal E} \sim {l_{D}^{2} \over t^{2}} \ K(t)^{2} \ .
\end{equation}
Dimensional arguments applied at this stage would then predict that
there exists a dynamical length scale $L_{s}(t) \sim l_{D}K(t)$, where
$l_{D} \sim (\nu t)^{1/2}$ is the diffusion scale.

In section IV we set out to establish this result by calculating the
two-point correlation function $C({\bf r},t)$. This task is
non-trivial, but we were able to extract the small and large scale
asymptotics of $C$. There were two unexpected results.  Firstly, the
small and large scale forms of $C$, although assuming a scaling form,
have different dynamic length scales. For the small distance scaling,
the dynamic length scale was found to be $L(t) \sim l_{D}/K(t)$, and
the dominant part of $C$ in this spatial regime is singular, and can
not be included in the scaling function. In terms of the
velocity-velocity correlation function $E({\bf r},t)$, this singular
piece is exactly ${\cal E}(t)$, with the scaling part of the
correlation function describing the non-local properties of $E$; i.e.
\begin{equation}
\label{velscaling1}
E({\bf r},t) \ \sim \ {\cal E}(t) \ + \ \left ({\nu \over l_{D}}\right
)^{2} {\tilde S} \left ( {{\bf r}\over L(t)} \right ) \ .
\end{equation}
The scaling function ${\tilde S}(x)$ has a power series expansion in
$x^{2}$, the first four coefficients of which may be inferred from
eq.(\ref{final}). The large distance scaling was found to be described
by the diffusive scale $l_{D}$ which confirmed {\it a posteriori} the
choice of this length scale in constructing the dynamic Reynolds
number. The leading term in this large distance regime was found to be
\begin{equation}
\label{finalform1con}
\delta C \sim \nu ^{2} K(t)^{(4-d^{2})/d} \exp[C_{d}''K(t)^{2(d+2)/d}]
\ \exp[-r^{2}/2l_{D}^{2}] \ ,
\end{equation}
The second unexpected result is that neither of the two dynamic length
scales coincide with the scale $L_{s}$ found from dimensional
considerations of ${\cal E}(t)$. This is explained in part by the fact
that the local energy decay is singular, and not contained in the
scaling form for $E$. This result is similar to the cases in critical
phenomena where caution is required in reading off scaling dimensions
from composite operators (like ${\bf v}^{2}$)\cite{amit}.  Generally,
these composite operators have their own scaling dimension, which may
be related to the scaling of a two-point correlation function only via
a small distance expansion (otherwise known as an operator product
expansion.)

In section V we introduced a mapping between BE and a toy model which
is well known from the field of disordered systems -- that being a
thermal particle in a harmonic background along with a random
potential. By considering simple scaling calculations on the toy
model, we were able to reproduce the previous result for ${\cal E}$ as
found in section III (up to prefactors). The scale $L_{s}$ appears
naturally in the toy model (as the typical root-mean-square
displacement of the particle), and it is interesting that this scale
is not a true dynamical scaling length, in that it may not be used to
collapse the two-point correlation function in a scaling plot.  It
would be interesting to understand this result more fully, either by
performing similar calculations to those described here for other
choices of initial distributions, or by calculating higher-order
correlation functions, to see if more exotic forms of scaling (like
multiscaling or intermittency) are present in these simple models.
The connection between the toy model and BE may be of some mutual aid
in both fields, at least in supplying new intuitive understanding of
these complementary problems.

\vspace{5mm}

The author would like to thank Sergei Esipov for important discussions
in the early stages of this investigation, and Alan Bray for an
enlightening conversation. The author acknowledges financial support
from the Engineering and Physical Sciences Research Council.

\newpage

\appendix

\section{}

In this appendix we outline the initial condition average over the
function
\begin{equation}
\label{psifapp}
\psi = \exp \left \lbrace -u \int d^{d}y \ g({\bf y},t) \exp \left [
\phi_{0}({\bf y})/2\nu \right ]\right \rbrace \ ,
\end{equation} 
where we have taken advantage of translational invariance and set
${\bf x}={\bf 0}$. (In fact, the translational invariance holds only
for times greater than $t_{0}$, since clearly the cellular initial
conditions allow only invariance under discrete transformations over a
period $l$.  Once the heat kernel has diffused beyond the cell scale,
the continuous translational invariance is recovered.)  The initial
distribution is as described in section II, namely, we divide space
into cells of volume $l^{d}$, and within each cell we assign the
velocity potential to be an independent random number drawn from a top
hat distribution of width $2\Phi $.  Explicitly we have
\begin{eqnarray}
\label{averpsi}
\nonumber \langle \psi \rangle & = & \int {\cal D}\phi _{0}({\bf
  y_{i}}) \ P[\phi _{0}] \psi [\phi _{0}] \\ & = & \prod \limits _{\bf
  Y} {1\over 2\Phi} \int \limits _{-\Phi}^{\Phi} d\phi_{0}({\bf Y}) \ 
\exp \left \lbrace -u \exp [ \phi_{0}({\bf Y})/2\nu ] \sum _{{\bf
    y_{i}}\in {\bf Y}}g({\bf y},t) \right \rbrace \ ,
\end{eqnarray}
where ${\bf Y}$ labels the cells, and ${\bf y}_{i}$ labels the
discretized points (on a scale of the lattice cut-off $a$) within a
given cell.  The integrals are easily performed in terms of the
exponential integral\cite{as}:
\begin{equation}
  {1\over 2\Phi} \int \limits _{-\Phi}^{\Phi} d\phi_{0} \exp \Bigl [
  -A \exp [\phi_{0}/2\nu ] \Bigr ] = {1\over K_{0}} \left [
  E_{1}(Ae^{-K_{0}/2}) - E_{1}(Ae^{K_{0}/2}) \right ] \ ,
\end{equation}
where $K_{0}=\Phi/\nu$ is the initial Reynolds number as defined in
the text.  To reach the result shown in the text, namely
eq.(\ref{psiav1}), two more steps are required. Firstly the above
result is re-exponentiated, so that the product over cells in
eq.(\ref{averpsi}) may be written in the exponent as a sum over cells.
Secondly, we use the fact that for times greater than $t_{0} =
l^{2}/\nu$, the heat kernel has smeared beyond the cell scale, so that
the sum over cells (on a scale $l$) and points within cells (on a
scale $a$) may be replaced once more by a continuum spatial integral.
In this way one finally arrives at the result given in the main text.

\section{}
This appendix is dedicated to the asymptotic evaluation of two
integrals that appear in the main text.

The first integral $L_{p}(u)$ is defined in eq.(\ref{intdef}), and is
used repeatedly in the present work. We have
\begin{equation}
\label{intdef1}
L_{p}(u) \equiv \int \limits _{0}^{\infty} ds \ s^{p}
[1-\exp(-ue^{-s})] \ .
\end{equation}
We require both the small and large-$u$ forms for this integral.  The
former is trivially obtained by expanding the integrand as a power
series in $u$. The asymptotic expansion of this integral for $u \gg 1$
is less simple.  For the precision required in the calculations, we
need both the dominant and sub-dominant terms. To extract these we
proceed as follows. We notice that for large-$u$, the second factor in
the integrand behaves very much like a step function centered at
$s=\ln (u)$. Thus, the dominant term will arise from replacing this
factor by $\theta (\ln(u)-s)$, and the sub-dominant term will arise
from finding the leading error made by this approximation.

So, explicitly we write
\begin{equation}
\label{step1}
L_{p}(u) = \int \limits _{0}^{\ln (u)} ds \ s^{p} + \int \limits
_{0}^{\infty} ds \ s^{p}T(s) \ ,
\end{equation}
where $T(s) = [1-\exp(-ue^{-s})]-\theta (\ln(u)-s)$.  The first term
gives the dominant contribution to the integral, which equals $[\ln
(u)]^{p+1}/(p+1)$.  To extract the main contribution from the second
term, we replace $s^{p}$ by $[\ln (u)]^{p}$ and perform the integral
over $T(s)$:
\begin{eqnarray}
\label{step2}
\nonumber \int \limits _{0}^{\infty} ds \ s^{p}T(s) & = & [\ln
(u)]^{p} \int \limits _{0}^{\infty} ds \ T(s) + O([\ln (u)]^{p-1}) \\ 
& = & [\ln(u)]^{p}\left \lbrace \int \limits _{\ln (u)}^{\infty} ds \ 
[1-\exp(-ue^{-s})] - \int \limits _{0}^{\ln (u)} ds \ \exp(-ue^{-s})
\right \rbrace + O([\ln (u)]^{p-1}) \ .
\end{eqnarray}
The first integral in the braces may be evaluated simply (using the
variable change $q=e^{-s}$) to give $\sum
_{n=1}^{\infty}(-1)^{n+1}/nn! \ $, whilst the same variable change
reduces the second integral to
\begin{equation}
\label{secondint}
\int \limits _{0}^{\ln (u)} ds \ \exp(-ue^{-s}) \ = \ \int \limits
_{1}^{u} {dq \over q} \ e^{-q} \ = \ E_{1}(1) + O(e^{-u}/u) \ .
\end{equation} 
Using the series expansion for the exponential integral\cite{as}, we
may combine the two results to give
\begin{equation}
\label{finalintegral}
\int \limits _{0}^{\infty} ds \ s^{p}T(s) = \gamma [\ln (u)]^{p} +
O([\ln (u)]^{p-1}) \ ,
\end{equation}
where $\gamma =0.57722...$ is Euler's constant.  This is the required
result for the subdominant terms.

The second integral appears in the evaluation of the large distance
scaling in section IV.  Referring to eq.(\ref{integ}) we need to
evaluate
\begin{equation}
\label{integ1}
J(K) = \int \limits _{c}^{\infty} du \ \exp \left \lbrace
-K(t)^{-(d+2)} [\ln (u)]^{d/2+1} \right \rbrace \ ,
\end{equation}
in the non-linear regime, $K(t) \gg 1$. The constant $c$ is a number
of order unity, and arises since we have used the large-$u$ form for
the integrand, and so we must cut off the integral at the lower end.
For notational convenience let us consider
\begin{equation}
\label{integ2}
M_{b}(N) = \int \limits _{1}^{\infty} du \ \exp \left \lbrace -(1/N)
[\ln (u)]^{b} \right \rbrace \ ,
\end{equation}
for $N \gg 1$, with $b>1$.  Then we can retrieve the integral of
interest from $J(K)=M_{d/2+1}(K^{d+2})$.

In order to cast the integral into a form suitable for steepest
descents, we make the variable change $x=N^{-1/(b-1)}\ln (u)$. We then
have
\begin{equation}
\label{stdec}
M_{b}(N) = N^{1/(b-1)} \int \limits _{0}^{\infty} dx \ \exp \left [
-N^{1/(b-1)} ( x^{b}-x ) \right ] \ .
\end{equation}
This integral is easily performed by steepest descents to give
(neglecting overall $b$-dependent constants):
\begin{equation}
\label{answer}
M_{b}(N) \sim N^{1/2(b-1)} \exp \left [ {(b-1)\over b} \left ( {N
  \over b} \right )^{1/(b-1)} \right ] \ ,
\end{equation}
from which one may retrieve the form given in eq.(\ref{finalform1}).

\section{}
In this appendix we give details of the manipulation of $I(u,v,{\bf
  R})$ into a series expansion for small and large $R$. As given in
eq.(\ref{ifn}) we have
\begin{eqnarray}
\label{ifn1}
\nonumber I(u,v,{\bf R}) = {2\over d\pi^{d/2}}\int d^{d}y & \ & \Biggl
\lbrace y^{2} - {{\bf y} \cdot {\bf R}\over 2} \left [ {ue^{-({\bf
      y}-{\bf R}/2)^{2}} - ve^{-({\bf y}+{\bf R}/2)^{2}} \over
  ue^{-({\bf y}-{\bf R}/2)^{2}} + ve^{-({\bf y}+{\bf R}/2)^{2}} }
\right ] \Biggr \rbrace
\\ \times & \ & \Biggl \lbrace 1 - \exp \left [ -ue^{-({\bf y}-{\bf
    R}/2)^{2}} -ve^{-({\bf y}+{\bf R}/2)^{2}} \right ] \Biggr \rbrace
\ .
\end{eqnarray}
As a first step, we expand the exponential term in the second factor
of the integrand to obtain
\begin{equation}
\label{blu}
I = \sum \limits _{n=1}^{\infty} {(-1)^{n+1} \over n!} \Lambda
_{n}(u,v,{\bf R}) \ ,
\end{equation}
where
\begin{eqnarray}
\label{blu1}
\nonumber \Lambda _{n}(u,v,{\bf R}) = {2\over d\pi^{d/2}}\int d^{d}y &
& \Bigl \lbrace y^{2} \left [ ue^{-({\bf y}-{\bf R}/2)^{2}}+
ve^{-({\bf y}+{\bf R}/2)^{2}} \right ]^{n} \\ 
\nonumber
& - & {{\bf y}\cdot {\bf
    R}\over 2} \left [ ue^{-({\bf y}-{\bf R}/2)^{2}}+ ve^{-({\bf
    y}+{\bf R}/2)^{2}} \right ]^{n-1} \left [ue^{-({\bf y}-{\bf
  R}/2)^{2}}-ve^{-({\bf y}+{\bf R}/2)^{2}} \right ] \Bigr \rbrace \ .\\
& & 
\end{eqnarray}
This cumbersome expression may be simplified greatly by expanding the
brackets as binomial series in powers of $u$ and $v$, and then
performing the gaussian integrals over ${\bf y}$. One is then left
with
\begin{equation}
\label{blu2}
\Lambda (u,v,{\bf R}) = n^{-(d/2+1)} \ \sum \limits _{m=0}^{n}
C^{n}_{m}u^{m}v^{n-m} \exp \left [ -{m\over n} \ (n-m)R^{2} \right ] \ 
.
\end{equation}
Combining eqs.(\ref{blu}) and (\ref{blu2}) reproduces the large-$R$
form given in eq.(\ref{larger}).

In order to cast $\Lambda $ into a small-$R$ form, we expand the
exponential terms in eq.(\ref{blu2}) and then binomially resum the
series in $u$ and $v$. This leaves us with
\begin{equation}
\label{blu3}
\Lambda (u,v,{\bf R}) = n^{-(d/2+1)} \ \sum \limits _{p=0}^{\infty}
{(-R^{2})^{p} \over n^{p} p!} \ (uv \partial _{u} \partial _{v})^{p} \ 
(u+v)^{n} \ .
\end{equation}
We now substitute this expression back into eq.(\ref{blu}) and make
the integral representation
\begin{equation}
\label{intrepres2}
n^{-(p+d/2+1)} = [\Gamma (p+d/2+1)]^{-1} \ \int \limits _{0}^{\infty}
ds \ s^{p+d/2} e^{-ns} \ .
\end{equation}
This allows the sum over $n$ to be performed as that for a geometric
series, and we are left with
\begin{equation}
  I(u,v,{\bf R}) = \sum \limits _{p=0}^{\infty} {(-R^{2})^{p} \over
    \Gamma (p+1) \Gamma (p+d/2+1)} (uv \partial _{u} \partial
  _{v})^{p}L_{p+d/2}(u+v) \ ,
\end{equation}
as given by eqs.(\ref{infsum}) and (\ref{funcfp}) in the main text.

\section{}

This appendix will give details of the rewriting of the function
$F_{p}(u,v)$ in moving from eq.(\ref{funcfp}) to (\ref{aaah}) in the
main text.  As given by eq.(\ref{funcfp}), we have defined
\begin{equation}
\label{funcfp11}
F_{p}(u,v) = {(-uv\partial _{u}\partial _{v})^{p} \over \Gamma (p+1)
  \Gamma (p+d/2+1)} \ L_{p+d/2}(u+v) \ .
\end{equation}
Since we are interested in the non-linear regime, we may expand the
integral $L$ as shown in Appendix B. Namely,
\begin{equation}
\label{expand}
L_{p+d/2}(u+v) = {\Delta ^{p+d/2+1} \over (p+d/2+1)} + \gamma \Delta
^{p+d/2} + O(\Delta ^{p+d/2-1}) \ ,
\end{equation}
where $\Delta \equiv \ln (u+v)$.  Then we define the quantities
$\chi_{p}(u,v;q) = (uv \partial_u \partial_v)^{p} \ \Delta ^{q}$ which
allow us to rewrite the above expression as
\begin{equation}
\label{aaahapp}
F_{p}(u,v) = {(-1)^{p} \over \Gamma (p+1) \Gamma (p+d/2+1)} \ \left [
{\chi_{p}(u,v;p+d/2+1) \over (p+d/2+1)} + \gamma \chi_{p}(u,v;p+d/2) +
O(\Delta ^{p+d/2-2}) \right ] \ ,
\end{equation}
which is of the form given in eq.(\ref{aaah}) in the main text.

The above steps are largely a matter of redefinition of various
quantities.  We must extract explicit forms for the functions
$\chi_{p}(u,v;q)$.  Clearly the zeroth function is just $\chi
_{0}(u,v;q) = \Delta ^{q}$. The subsequent functions may be written as
\begin{eqnarray}
\label{aaah2}
\nonumber \chi_{p}(u,v;q) & = & (uv \partial_u \partial_v)^{p} \ 
\Delta ^{q} \\ & = & f_{p}(u,v;q)\Delta ^{q-1} + g_{p}(u,v;q) \Delta
^{q-2} + O(\Delta ^{q-3}) \ .
\end{eqnarray}
In order to unambiguously determine the scaling properties of the
correlation function, it is sufficient to explicitly evaluate $C$ to
$O(R^6)$. However, given the singular nature of the scaling in this
problem, we shall proceed to calculate the $O(r^{8})$ terms as well,
as a useful check. This in turn necessitates calculating the
coefficients $f_{p}$ and $g_{p}$ for $p=1,2,3,4$. With the application
of brute force algebra, we obtain:
\begin{eqnarray}
\label{ffuncs}
\nonumber f_{1}(u,v;q) = & - & q {uv \over (u+v)^{2}} \\ \nonumber
f_{2}(u,v;q) = & + & q {uv \over (u+v)^{4}} \ (u^{2}-4uv+v^{2}) \\ 
\nonumber f_{3}(u,v;q) = & - & q {uv \over (u+v)^{6}} \ 
(u^{4}-26u^{3}v+66u^{2}v^{2} -26uv^{3}+v^{4}) \\ \nonumber
f_{4}(u,v;q) = & + & q {uv \over (u+v)^{8}} \ 
(u^{6}-120u^{5}v+1191u^{4}v^{2}
-2416u^{3}v^{3}+1191u^{2}v^{4}-120uv^{5}+v^{6}) \ , \\ & &
\end{eqnarray}
and
\begin{eqnarray}
\label{gfuncs}
\nonumber g_{1}(u,v;q) = & - & (q-1)f_{1} \\ \nonumber g_{2}(u,v;q) =
& - & 2(q-1)f_{2}-q(q-1){u^{2}v^{2} \over (u+v)^{4}} \\ \nonumber
g_{3}(u,v;q) = & - & 3(q-1)f_{3}+q(q-1){u^{2}v^{2} \over (u+v)^{6}} \ 
(17u^{2}-52uv+17v^{2}) \\ \nonumber g_{4}(u,v;q) = & - & 4(q-1)f_{4}
 - q(q-1){u^{2}v^{2} \over (u+v)^{8}} \ 
(129u^{4}-1648u^{3}v+3538u^{2}v^{2} -1648uv^{3}+129v^{4}) \ . \\ & &
\end{eqnarray}

\section{}

In this final appendix we give a brief description of the final steps
required in order to perform the $u$ and $v$ integrals, so as to
derive the final form of the correlation function given in
eq.(\ref{final}).

Referring to the main text, we see that combining eqs.(\ref{corrfn}),
(\ref{psifn}) and (\ref{infsum}), we may write the correlation
function in the form
\begin{equation}
\label{almost}
C({\bf r},t)=8\nu^{2} \int \limits _{c}^{\infty} {du \over u} \int
\limits _{c}^{\infty} {dv \over v} \ \exp [-\epsilon F_{0}(u,v)] \left
\lbrace 1 - \exp \left [ -\epsilon \sum \limits _{p=1}^{\infty} R^{2p}
F_{p}(u,v) \right ] \right \rbrace \ ,
\end{equation}
where $c$ is a number of order unity, required simply to cut off the
integrals at their lower limits, given we are using the asymptotic
form for the integrand in the non-linear regime. The remaining steps
are easy to describe although rather tedious to perform in practice.
We expand the exponential in the last factor of the above integrand in
powers of $R$, and then integrate over $u$ and $v$. We shall
explicitly demonstrate this for the dominant part of the $O(R^{2})$
term. Using the asymptotic forms of the functions $F_{0}$ and $F_{1}$
from Appendix D, we have
\begin{equation}
\label{leading1}
O(R^{2}) \sim {-8\nu ^{2}\epsilon\over(d/2+2)\Gamma (d/2+2)} \ \int
\limits _{c}^{\infty} {du\over u} \int \limits _{c}^{\infty} {dv \over
  v} \ f_{1}(u,v;2+d/2) \Delta ^{d/2+1}\exp \left [ -{\epsilon \over
  \Gamma (d/2+2)}\Delta ^{d/2+1} \right ] \ .
\end{equation}
Remembering that $\epsilon = \Gamma(d/2+2)/K(t)^{d+2}$, and using the
form of $f_{1}$ from the previous appendix, we may rewrite this term
as
\begin{equation}
\label{leading2}
O(R^{2}) \ {\rm term} \ \sim 8\nu ^{2}K(t)^{-(d+2)} \ \int \limits
_{c}^{\infty} du \int \limits _{c}^{\infty} dv \ (u+v)^{-2} \Delta
^{d/2+1}\exp \left [ -K(t)^{-(d+2)}\Delta ^{d/2+1} \right ] \ .
\end{equation}
Now the double integral has the form
\begin{equation}
\label{intbyparts}
\int \limits _{c}^{\infty} du \int \limits _{c}^{\infty} dv \ 
(u+v)^{-2} A(u+v) = \int \limits _{c}^{\infty} du \int \limits
_{u}^{\infty} dw \ w^{-2} A(w) \simeq \int \limits _{c}^{\infty} {du
  \over u} \ A(u) \ ,
\end{equation}
where the last step was achieved using integration by parts on $u$.
The final integral may be easily performed by substituting $x=\ln(u)$,
thus yielding the first term in eq.(\ref{final}).

The integrals required at higher orders in $R^{2}$ may all be
performed by changing variables from $(u,v)$ to $(u,w=u+v)$ and
performing the appropriate number of integrations by parts on $u$. In
general the `boundary' terms do not vanish, but they are negligible in
the limit of interest, namely $K(t) \gg 1$. (This is because the
general form of the function $A(w)$, which appears as a spectator in
these integral manipulations, is proportional to $\exp
[-K(t)^{-(d+2)}[\ln (w)] ^{d/2+1}]$, which decays faster than any
power.) A final point worth mentioning is that the sub-dominant terms
(i.e. those terms involving the coefficients $g_{p}$) are required
since the $(u,v)$ integrals over the leading terms containing $f_{p}$
all vanish for $p>1$. This leads to the unusual scaling form described
in the text, in which the dominant term may not be cast as part of the
scaling function; and explains our cautionary calculation of the
$O(R^{8})$ term, which indeed confirms this singular scaling, i.e.
the integrals over $f_{4}$ vanish exactly.


\begin{references}

\bibitem{b1} J. M. Burgers, Proc. Acad. Sci. Amsterdam {\bf 43}, 2 (1940).

\bibitem{karp} V. I. Karpman, {\it Non-linear Waves in Dispersive Media}
(Pergamon, Oxford, 1975).

\bibitem{kdv} D. J. Korteweg and G. de Vries, Phil. Mag. {\bf 39}, 
422 (1895).

\bibitem{kpz} M. Kardar et al., Phys. Rev. Lett.
  {\bf 56}, 889 (1986).

\bibitem{abr} D. B. Abraham, Phys. Rev. Lett. {\bf 44}, 1165 (1980);
  G. Forgacs et al., Phys. Rev. Lett.  {\bf 57}, 2184 (1986); B. Derrida
  et al., J. Stat. Phys.  {\bf 66}, 1189 (1992).

\bibitem{huse} D. A. Huse and C. L. Henley, Phys. Rev. Lett. {\bf 54},
  2708 (1985).

\bibitem{blat} G. Blatter et al., Rev. Mod. Phys. {\bf 66}, 1125
  (1994).

\bibitem{gurb} S. N. Gurbatov et al., {\it Non-linear Random Waves and 
Turbulence in Nondispersive Media} (Manchester University Press, Manchester,
1991).

\bibitem{lss} M. Vergassola et al., Astro. Astrophys. {\bf 289}, 325 (1994).

\bibitem{amit} D. J. Amit, {\it Field Theory, the Renormalization Group and 
Critical Phenomena} 2nd Edition (McGraw-Hill, New York, 1984). 

\bibitem{bray} A. J. Bray, Adv. Phys. {\bf 43}, 357 (1994).

\bibitem{glau} R. J. Glauber, J. Math. Phys. {\bf 4}, 294 (1963).

\bibitem{brnew} T. J. Newman and A. J. Bray, J. Phys. A {\bf 23}, L279 
(1990). 

\bibitem{b2} J. M. Burgers, {\it The Non-linear Diffusion Equation}
  (Reidel, Boston, 1974).

\bibitem{kida} S. Kida, J. Fluid Mech. {\bf 93}, 337 (1979).

\bibitem{esnew} S. E. Esipov and T. J. Newman, Phys. Rev. E {\bf 48}, 1046
(1993).

\bibitem{es2} S. E. Esipov, Phys. Rev. E {\bf 49}, 2070 (1994).

\bibitem{gurb1} S. N. Gurbatov and A. I. Saichev, Sov. Phys. JETP {\bf 53},  
347 (1981).

\bibitem{sinai} Ya. G. Sinai, Comm. Math. Phys. {\bf 148}, 601 (1991).

\bibitem{jeff} H. Jeffreys and B. Jeffreys, {\it Methods of Mathematical
Physics} 3rd Edition (Cambridge University Press, Cambridge, 1980).

\bibitem{hopf} E. Hopf, Comm. Pure Appl. Math. {\bf 3}, 201 (1950).

\bibitem{cole} J. D. Cole, Quart. Appl. Math. {\bf 9}, 225 (1951).

\bibitem{derr} B. Derrida, Phys. Rev. B {\bf 24}, 2613 (1981).

\bibitem{new} T. J. Newman, Phys. Rev. E {\bf 49}, R2525 (1994) {\rm and}
Phys. Rev. E {\bf 51}, 4212 (1995). 

\bibitem{as} {\it Handbook of Mathematical Functions} 10th Edition, 
M. Abramowitz and I. A. Stegun Eds. (Dover, New York, 1972).

\bibitem{tm} J. Villain et al., J. Phys. C {\bf 16}, 2588 (1983); 
A. Engel, J. Physique Lett. {\bf 46}, L409 (1985); J. Villain,
J. Phys. A {\bf 21}, L1099 (1988); M. M\'ezard and G. Parisi, J. Phys. I
France {\bf 2}, 2231 (1992).

\bibitem{bray1} I am indebted to A. J. Bray for suggesting the scaling
arguments in this section.

\bibitem{mez} V. Dotzenko and M. M\'ezard, cond-mat/9611017.

\end{references}
\end{document}